\documentclass[12pt]{article}
\usepackage{amssymb,amsmath,amsfonts,amsthm,graphicx}
\def\be{\begin{equation}}
\def\ee{\end{equation}}

\begin{document}

\begin{center}
{\Large\bf The JWST and standard cosmology}
\vspace{.3in}
\\{\bf A A Coley},
\\Department of Mathematics \& Statistics, Dalhousie University,\\
Halifax, Nova Scotia, Canada B3H 3J5
\\Email: alan.coley@dal.ca
\vspace{.1in}

\end{center}

\begin{abstract}

Recent observations from the 
James Webb Space Telescope have identified a population of massive galaxy sources ($\mathrm{>10^{10}\ M_\odot}$) at $z>7-10$, 
formed less than 700 Myr after the Big Bang. 
Such massive galaxies do not have
enough time to form within the standard 
cosmological model, and hence these
observations significantly  challenge standard
cosmology.
A number of possible solutions to this problem have been put forward, including
an enhancement of the gravitational force in a modified theory of gravity and
the claim that massive primordial black holes, which were created  in the 
early universe before galaxy formation, might
provide seeds for galaxies and black holes to subsequently form.
We discuss two more exotic possibilities. Black holes can persist through a cosmological bounce 
and constitute large seeds formed in the previous cosmic cycle 
existing before current galaxy formation. And spikes, 
both incomplete spikes that occur in the early initial
oscillatory regime of general cosmological models and permanent spikes that can form in inhomogeneous models
at later times, 
could provide a  mechanism for generating large structures early in the Universe.


\end{abstract}

\newpage

\section{Introduction}

Cosmology is the investigation of the large scale behaviour of the Universe when small-scale features can be neglected \cite{coleyellis}. The cosmological principle asserts that on the largest scales the Universe can be accurately modeled by the Friedmann-Lemaıtre-Robertson-Walker (FLRW) solution of the Einstein Field Eqns. (EFE), which is both spatially homogeneous and isotropic. The
concordance $\Lambda$CDM FLRW cosmology (or simply standard cosmology for short) has a cosmological constant, $\Lambda$, and contains cold dark matter.
Inflation is often assumed to be the mechanism for the generation
of almost scale invariant fluctuations that
act as seeds for the subsequent formation of large scale structure (however,
possible alternatives include
recurring spikes \cite{art:ColeyLim2012} and fluctuations generated before a possible cosmological bounce \cite{steinhardt}).
Early universe inflation is usually taken to be part of the standard model, and spatial curvature is assumed to be zero or vanishingly small.

To date, standard cosmology has been extremely
successful in explaining all observations up to various 
anomalies and tensions \cite{Buchert}. 
In particular,
the Hubble Space Telescope (HST) determination of the local value of the Hubble constant based on
direct measurements of supernovae \cite{Riess}
is now statistically significantly higher than the value derived from 
CMB data from the Planck satellite in the standard model  \cite{PLANCK}.
We also note that the standard model does require, as yet undetected, sources of a  current dynamically dominant dark
energy density.

One of the greatest challenges in cosmology is understanding the origin and growth of  structure in the universe. 
The CMB anisotropies and structure observed on large angular scales are computed using linear perturbations about the standard cosmological background
model (which is usually assumed to follow a period of inflation).
Indeed, in the standard cosmology it is often assumed that cosmic structure at sufficiently
large scales grew out of small initial fluctuations at early times, whose evolution can be studied within (cosmological) linear perturbation theory (LPT) \cite{Durrer}.
At late times and sufficiently small scales, fluctuations of the cosmic density
are no longer small, and LPT is not adequate
to study structure formation on scales of a few Mpc and less. It
is subsequently necessary to treat clustering non-linearly using N-body computations.

\section{JWST}

In standard cosmology spatial homogeneity is only statistically valid
on scales larger than approximately 100-115 Mpc, and there can only exist spatial inhomogeneities up to scales of about 100 Mpc.
But the observed Universe has web-like features dominated
by enormous voids within bubble walls, and with filamentary structures containing clusters of galaxies. In fact, currently the distribution of matter is certainly not spatially homogeneous
on scales less than 150-300 Mpc.
Indeed, the largest structures  observed to date include
the  Sloan Great Wall, with a 400 Mpc long  filament, and 
Large Quasar Groups which are characteristically of size 70-350 Mpc.
and perhaps as large as $\sim$ 500 Mpc. (and at a redshift  $\sim 1.27$) \cite{art:Clowes2012}.
In addition, there are two enormous local voids approximately 35 to 70 Mpc wide associated
with the so-called velocity anomaly, and there is observational evidence for a void
as large as 430 Mpc. and for void complexes on scales of up
to 450 Mpc. 
Such large spatial inhomogeneities in the distribution
of superclusters and voids on scales of 200-300 Mpc and above, and as observed by the HST~\cite{hst1},
are potentially inconsistent
with  the standard model
\cite{art:ShethDiaferio2011}. 

Indeed, El Gordo, the
most massive galaxy  cluster ever observed, formed far too early in the evolution of the Universe and with too large a mass and collision speed to be compatible with the standard model. In addition,
the emergence of super-massive black holes (SMBHs) early in the universe at redshifts z $\ge 7$ 
in active galactic nuclei and quasars span a range extending up to masses of nearly $10^{11}M_{\odot}$~\cite{McConnell2011}, and are in conflict with our theoretical explanation of the accretion of BH seeds and their potential rapid growth in the first 800 million years after the Big Bang \cite{2018Natur.553}.

Therefore, within the currently accepted theory of hierarchical structure formation in standard $\Lambda$CDM cosmology, it is difficult to understand how galaxies could accumulate so much mass through mergers or accretion alone, and predict that the first
galaxies should only appear between redshifts of approximately $15 \leq z \leq 20$  \cite{Mason2023}. This raises the question of whether there is sufficient time for these objects to grow to a BH with  mass $\sim 10^9-10^{10}M_{\odot}$ in such a short time after the Big Bang.

All of these issues are exarcebated by
recent observations from the 
James Webb Space Telescope (JWST)  \cite{JWST}, which
have detected a large number of ultra-high-redshift luminous galaxy candidates beyond $z>10$,
and some up to $z>15$ \cite{jwst-1}, revealing new galaxies undetected even in the deepest HST observations \cite{nircam_performance} 
(and previously unseen in HST due to the Lyman-$\alpha$ break caused by the neutral intergalactic medium). That is, JWST has identified the emergence of a population of (unusually) massive galaxy sources ($\mathrm{>10^{10}\ M_\odot}$) at $z>7-10$ ~\cite{liu}, less than 700 Myr after the Big Bang \cite{imp-gal}, with the
most distant galaxy being Gs-13, observed just 350 Myr after the Big Bang.
[Further analysis is needed to check the nature of these new  high redshift objects;
while some galaxies have already been confirmed spectroscopically, there is mounting evidence that a sub-sample of the candidates with particularly red inferred UV colors are, in fact, lower redshift contaminant galaxies at $z\sim2-6$.]

In addition, current JWST observations  have detected a massive BH with $M_{BH} \sim 10^7 - 10^8 M_{\odot}$ at $z \sim 10.3$ \cite{2023arXiv230515458B} and  high-redshift quasars, which reveal that many SMBHs were in place less than 700 Million years after the Big Bang. 
Therefore, JWST has led to the surprising conclusion  that the early universe, younger 
than one billion years, is densely populated by well developed galaxies and quasars (SMBHs)
that should not have had time to form within the standard 
$\Lambda$CDM cosmology.
The Big Bang is still present, but it no long necessarily occurs at a 'beginning'.
Distant galaxies also contain heavy elements (heavier than helium) not consistent with theory.
{\em These  striking observations are incompatible with, and therefore
significantly  challenge, the standard
cosmological scenario} \cite{Foroconi}.

Researchers have investigated a number of solutions to the problem of the emergence of SMBHs when the Universe was just $\sim 800~ Myr$, grouped roughly into  small seed scenarios \cite{2019MNRAS.484.2575T}, 
which encompass mechanisms where the initial mass of the BH is relatively low
$\sim$ 100 $M_{\odot}$ but rapid growth happens,
and massive seed scenarios \cite{2019MNRAS.486.3892R} with an initial mass greater than $\sim$ 10,000 $M_{\odot}$ but a smaller growth rate.

The key to rapid SMBH growth is exceptionally high rates of accretion in which the BH continuously gathers mass from its surrounding environment.  There were certainly regions in the early universe with sufficient gas to provide the necessary fuel for accretion.
It is usually  taken that SMBHs that occur
in centres of large galaxies are created by matter accretion
to the over-density at the galactic centre, 
but the approximate time necessary is longer than the age of the current universe
of about 15 billion years, and much longer than the considerably 
younger universe at $z\sim 10$.

It might still be possible to reconcile observations with theory \cite{Mason2023}. One possibility is that these sources are actually active galactic nuclei \cite{imp-gal}.
A more popular solution to the aforementioned growth problem might be via an enhancement of the gravitational force. Some authors achieve this via dark matter halos \cite{2023arXiv231006898S}, but others prefer to try and solve the problem by utilizing an alternative theory of gravity to general relativity (GR) (or possibly backreaction effects on large scales \cite{coleyellis}); in particular, it has been argued that the Modified Gravity (MOG)  \cite{Moffat} can explain how SHBHs can accumulate such a huge mass at redshifts $z \ge 7$ \cite{Moffat2006}.

Perhaps more conventionally it has been claimed that massive primordial black holes (PBH), which were created first in the 
early universe at a pre-stellar epoch  \cite{Carr}, could 
seed  galaxy and quasar formation in the very young universe
~\cite{DS}.
Indeed, the observed massive galaxy candidates can be explained,
contrary to the  $\Lambda\rm CDM$ model, if structure formation is sped up by
massive ($\gtrsim {10^{9}\ \rm M_{\odot}}$) PBHs 
that enhance primordial density perturbations.
The model also predicts an earlier emergence of galaxies, seeded in part by PBH \cite{DS}. A population of red candidate
massive galaxies ($ > 10^{10} M_\odot$) at $ 7.4 \lesssim z \lesssim 9.1$, 500–700 Myr after the Big Bang, 
including one galaxy with a possible mass of $\sim 10^{11} M_\odot$ which is far too massive
to have been created so early in the universe, have been observed \cite{imp-gal}.  
Let us also consider 2 more exotic possibilities.


\section{Persistence}

Gravitational dynamics can classically
lead to the initial cosmological (or Big Bang) singularity being replaced by a
Big Bounce, a smooth transition from contraction to an expanding universe \cite{brand}.
Indeed, in many cosmological settings, the present phase of  universe expansion would have been preceded by a collapsing phase. The  bounce, which necessarily connects these two separate phases and which necessarily violates the null energy condition, could arise from: (i) classical effects associated with, for example, a cosmological constant \cite{lemaitre}, a scalar field \cite{steinhardt} or from another modified theory of gravity \cite{tors1}
in which the
bounce may occur below the Planck density, or (ii)
quantum gravitational effects perhaps associated with string
theory \cite{Turok} or loop quantum gravity \cite{qgt} in which the
bounce presumably occures at the Planck density.

It is of interest to investigate \cite{cc} whether BHs can persist through a cosmological bounce and what possible effects they might have on the large-scale dynamics.
Such BHs can be divided into two classes: (i) ``pre-crunch black holes''  that persist in a universe that recollapses to a big crunch and subsequently evolves into a new expansion phase; and (ii) ``big-crunch black holes'' that are actually produced by the high matter density at the bounce itself. We shall utilize the term ``pre-big-bang black hole'' (PBBBH) to cover both of these cases. PBBBHs of class (i) could be very massive, similar to those that form as a result of stellar collapse or occur at the centres of galactic nuclei in the present universe. PBBBHs of class (ii) could be very small and emit quantum radiation, but they would likely complete their evaporation {\it after} the bounce and so might be observationally indistinguishable from  the PBHs which formed after the Big Bang.

Recently, some exact dynamical solutions  were derived  \cite{cc} which describe a regular hyperspherical lattice of black holes in a cosmological background dominated by a scalar field at the bounce. 
More precisely, an exact solution of the Einstein constraint equations was presented which can
be used to develop a 4-dimensional dynamical solution  
in which multiple distinct black holes propagate through the bounce. 
These results 
demonstrate that there certainly do exist exact solutions in which multiple BHs do
not merge before nor at the bounce, so that consequently the BH can indeed
persist through a bounce. 
Since the bounce can occur well below the 
Planck density, such a classical approach is legitimate.

There are some important cosmological consequences of this potential persistence. In different mass ranges PBBBHs could contribute to the dark matter, provide seeds for galaxies, generate entropy and even drive the bounce itself \cite{essay}.
If PBBBHs are randomly distributed in more general and more physically realistic non-symmetric
collapse it would be expected that a process of hierarchical and partial merging would occur, in which progressively larger horizons would form around groups of BHs, with the result that the characteristic BH mass would steadily increase.

The possibility  that the dark matter could comprise PBHs has recently gained popularity. However, PBBBHs could be equally plausible and an interesting question would then be whether 
there is any~observational
way of distinguishing PBBBHs from the more conventional PBHs.

Of more relevance here, it
is believed that most galactic nuclei  contain SMBHs, extending from $10^{6}M_\odot$ to $10^{10}M_\odot$ and which are already in place by a redshift of $\sim 10$. As noted above,
it is difficult to account for how such enormous BHs might have formed so early unless perhaps there were large seed BHs already existing before galaxy formation
\cite{DS}. 
The suggestion that these seeds might be PBHs forming soon after the Big Bang has consequently gained traction. However, of course, the seeds could equally well be PBBBHs,  left over from the galaxies formed in the previous cosmic cycle. It is also possible that the galaxies themselves might not survive the bounce but the central SMBHs may do so.

There are several areas in which the current analysis could be extended. 
One feature of the solutions studied to date is that the number of BHs ($N$) is finite and relatively small. This, of course, does not resemble the situation in the  real universe, where the number of SMBHs is of order $10^{10}$  and the number of intermediate mass BHs could be of order $10^{20}$. However, $N$ could be arbitrarily large in a flat or open model.

\section{Spikes}

Spikes were first observed to occur in the context of vacuum orthogonally transitive $G_2$ models~\cite{art:BergerMoncrief1993}, which represent a dynamical but spatially inhomogeneous gravitational 
distortion.   Lim subsequently discovered an exact solution for spikes~\cite{art:Lim2008}.
In \cite{art:ColeyLim2012}
we explicitly demonstrated that spikes naturally occur in a class of
non-vacuum $G_2$ cosmologies and, due to
gravitational instability, leave small residual imprints in the
form of matter perturbations.
We have especially been interested in studying recurring and complete
spikes formed in the oscillatory regime (or recurring spikes for short)~\cite{art:Lim2008,art:Limetal2009}, and their subsequent
imprint on matter and structure formation.

The residual matter overdensities produced from recurring spikes
form on surfaces, not at isolated points.
In general, two spikes surfaces may intersect along a curve,
and this curve may intersect with a third spike surface at a point, consequently leading to matter inhomogeneities being formed on a web of surfaces, curves and points.
Therefore, it is plausible that
filamentary structures and voids would generally evolve in this scenario.

With the inclusion of a tilted fluid, there is a second mechanism in which matter
inhomogeneities are generated because of the non-negligible 
divergence terms caused by an instability in the tilt ~\cite{art:ColeyLim2012}.
We studied the effect of a sign change in the tilt numerically, which 
leads to  a large divergence creating both overdensities and underdensities. Indeed,
we found that it is the tilt instability that may play the dominant role in the formation of local matter inhomogeneities in these models.

So far we have discussed how spikes generate matter overdensities in a radiation fluid in a
special class of inhomogeneous models. These spikes occur in the initial
oscillatory regime of general cosmological models.  
Spike solutions support the BKL conjecture \cite{BKL}, which asserts
that within GR and for a generic inhomogeneous cosmology the approach to the
(past) spacelike singularity is (generally) vacuum dominated, local, and oscillatory,
except at (isolated) spike points.

Spikes can also occur in other regimes.
The heuristic  dynamical reason for spike formation is that the initial data straddle the stable manifold of a saddle point in the dynamical state space. 
As the solution evolves, the data on either side of the stable manifold begin to approach the saddle point and then subsequently evolve away from the local neighbourhood of the saddle point with the result that neighbouring worldlines diverge from each other, thereby
producing a spiky profile at later times. 
The EFE admit a number of self-similar solutions, 
which are represented as equilibrium points in the reduced expansion-normalized state space~\cite{book:WainwrightEllis1997}, many of which are saddle points.
It is consequently natural to anticipate that spikes will occur around these saddle points, or more general unstable subspaces. In general, such spikes may occur on sets of different dimensions (i.e., points, curves or surfaces).

We have demonstrated that permanent spikes occur in (spherically symmetric dust) Lema\^{\i}tre-Tolman-Bondi models (LTB) models  \cite{LTB}, selected since the solutions are exact
and can be studied in a straightforward manner using qualitative dynamical systems methods. 
Indeed, the LTB cosmologies were shown to have permanent spikes form (around the vanishing $E(r)$ worldlines).
We anticipate that similar permanent spikes will occur at the boundary between ever-expanding and recollapsing regions, and other unstable boundaries, in more general models.
Both the incomplete spikes~\cite{art:ColeyLim2012}, and especially the late time non-generic inhomogeneous spikes  \cite{LTB}, might provide a  mechanism within GR for generating
exceptionally large structures observed in the Universe.

\newpage

\section*{Acknowledgement}

AAC is supported by NSERC.

\end{document}